\documentclass[%
 reprint,
superscriptaddress,
nofootinbib,
 amsmath,amssymb,
 aps,
prx,
]{revtex4-2}
\usepackage[dvipsnames]{xcolor}
\usepackage{graphicx}
\usepackage{dcolumn}
\usepackage{bm}
\usepackage{hyperref}

\usepackage{braket}
\usepackage[utf8]{inputenc} 
\usepackage{mathtools}
\usepackage[bottom]{footmisc}
\usepackage{hyperref}
\usepackage{amsfonts}
\usepackage{amsmath}
\usepackage{slashed}
\usepackage{amsthm}
\usepackage[normalem]{ulem}
\usepackage{soul}

\begin{document}

\preprint{APS/123-QED}

\title{Atomic Scale Quantum Anomalous Hall Effect in Monolayer Graphene/$\rm MnBi_{2}Te_{4}$ Heterostructure}

\author{Yueh-Ting Yao}
\affiliation{Department of Physics,\;National Cheng Kung University,\;Tainan,\;70101,\;Taiwan}

\author{Suyang Xu}
\email{suyangxu@fas.harvard.edu}
\affiliation{Department of Chemistry and Chemical Biology,\;Harvard University,\;Cambridge,\;MA,\;02138,\;USA}

\author{Tay-Rong Chang}
\email{u32trc00@phys.ncku.edu.tw}
\affiliation{Department of Physics,\;National Cheng Kung University,\;Tainan,\;70101,\;Taiwan}
\affiliation{Center for Quantum Frontiers of Research and Technology (QFort),\;Tainan,\;70101,\;Taiwan}
\affiliation{Physics Division, National Center for Theoretical Sciences,\;Taipei,\;10617,\;Taiwan}

\date{\today} 

\begin{abstract}
The two-dimensional quantum anomalous Hall (QAH) effect is direct evidence of non-trivial Berry curvature topology in condensed matter physics. Searching for QAH in 2D materials, particularly with simplified fabrication methods, poses a significant challenge in future applications. Despite numerous theoretical works proposed for the QAH effect with $C=2$ in graphene, neglecting magnetism sources such as proper substrate effects remain experimental evidence absent. In this work, we propose the QAH effect in graphene/$\rm MnBi_{2}Te_{4}$ (MBT) heterostructure based on density-functional theory (DFT). The monolayer MBT introduces spin-orbital coupling, Zeeman exchange field, and Kekul$\rm \acute{e}$ distortion as a substrate effect into graphene, resulting in QAH with $C=1$ in the heterostructure. Our effective Hamiltonian further presents a rich phase diagram that has not been studied previously. Our work provides a new and practical way to explore the QAH effect in monolayer graphene and the magnetic topological phases by the flexibility of MBT family materials.
\end{abstract}

\maketitle



\section{\label{sec:level1}Introduction}

\par Searching the non-trivial topological states in real material is a primary goal in modern condensed matter physics\cite{NO1,NO2}. The quantum anomalous Hall(QAH) effect, which is the experimental observation of the Chern insulator, shows promising potential in electronic devices with low power consumption\cite{NO3}. Until now, the QAH effect has been observed in magnetically doped topological insulators\cite{NO4,NO5,NO6}, the intrinsic magnetic topological insulator $\rm MnBi_{2}Te_{4}$ families\cite{NO7,NO8}, and twisted multilayer moir$\rm \acute{e}$ systems\cite{NO9,NO10}. The exploration of QAH materials continues to attract considerable attention in applications.

\par Graphene, as the prototypical 2D material, boasts numerous advantages, including remarkable robustness, high electronic mobility, and a massless Dirac band structure. In particular, the massless Dirac band structure can be viewed as a critical point adjacent to various topological phases. For instance, the introduction of spin-orbit coupling(SOC) transforms monolayer graphene into a quantum spin Hall(QSH) insulator\cite{NO11,NO12}. Moreover, with the incorporation of both SOC and Zeeman exchange field, monolayer graphene can evolve into a quantum anomalous Hall insulator\cite{NO13,NO14}. Despite numerous theoretical works proposed the QAH effect in graphene\cite{NO13,NO14,NO15,NO16,NO17,NO18,NO19}, experimental evidence remains absent. This discrepancy may stem from prior investigations neglecting magnetism sources such as proper substrate effects. In recent years, the intrinsic topological insulator $\rm MnBi_{2}Te_{4}$(MBT) families have emerged as a promising platform for exploring magnetic topological phases\cite{NO20}. The monolayer MBT can be viewed as intercalating a Mn-Te bilayer into the center of a $\rm Bi_2Te_3$ quintuple layer, forming a Te-Bi-Te-Mn-Te-Bi-Te septuple layer. Notably, the MBT lattice exhibits an almost perfect lattice match with the $\sqrt{3}$-by-$\sqrt{3}$ supercell of monolayer graphene. Indeed, researchers have investigated the potential use of the $\rm Bi_2Te_3$ families\cite{NO21} (which have the same lattice constant as MBT but lacking magnetism) to introduce SOC in graphene, both theoretically and experimentally\cite{NO22,NO23,NO24}. Drawing inspiration from these findings and preceding studies, our present work explores the innovative prospect of utilizing MBT as a substrate for monolayer graphene. This approach aims to concurrently introduce SOC, Kekul$\rm \acute{e}$ distortion, and Zeeman exchange field, thereby inducing the QAH effect in monolayer graphene.

\par In this paper, we analyze the electronic structure and present a comprehensive topological phase diagram for the monolayer graphene/MBT heterostructure. Our result reveals that the proximity to monolayer MBT opens a band gap in graphene by introducing SOC and/or Kekul$\rm \acute{e}$ distortion. Specifically, SOC causes graphene to transition from a Dirac semimetal to the QSH state\cite{NO23}, while the Kekul$\rm \acute{e}$ distortion introduces a superlattice potential, mixing different valley states and transforming graphene into a normal insulator(NI)\cite{NO23,NO24,NO25}. The interplay of these gap-opening mechanisms and the Zeeman exchange field from the monolayer MBT significantly enriches the topological phase diagram, giving rise to spin Chern insulator (observed QSH effect), Chern insulator (observed QAH effect), two-fold/four-fold/eight-fold degenerate Dirac semimetal, and NI phases. The conventional understanding for inducing QAH effect in graphene posits that each valley contributes a unit conductance $e^2/h$, leading to a quantized Hall conductivity at $\sigma_{xy}=2e^2/h$, without considering the source of Zeeman exchange field\cite{NO13,NO14,NO15,NO16}. In contrast, incorporating the Zeeman exchange field from the MBT substrate, coupled with valley mixing from Kekul$\rm \acute{e}$ distortion, results in a quantized Hall conductivity at $\sigma_{xy}=e^2/h$, representing the ground state of Chern insulator with $C=1$. Our first-principles calculations substantiate the existence of the $C=1$ ground state with $\sigma_{xy}=e^2/h$ in graphene/MBT heterostructure. The emergence of Kekul$\rm \acute{e}$ distortion at the graphene/MBT interface presents a unique opportunity for achieving tunable Chern numbers in the heterostructure system.

\section{\label{sec:level1}Results}

\par
\begin{figure}[t]
\centering
\includegraphics[width=\linewidth]{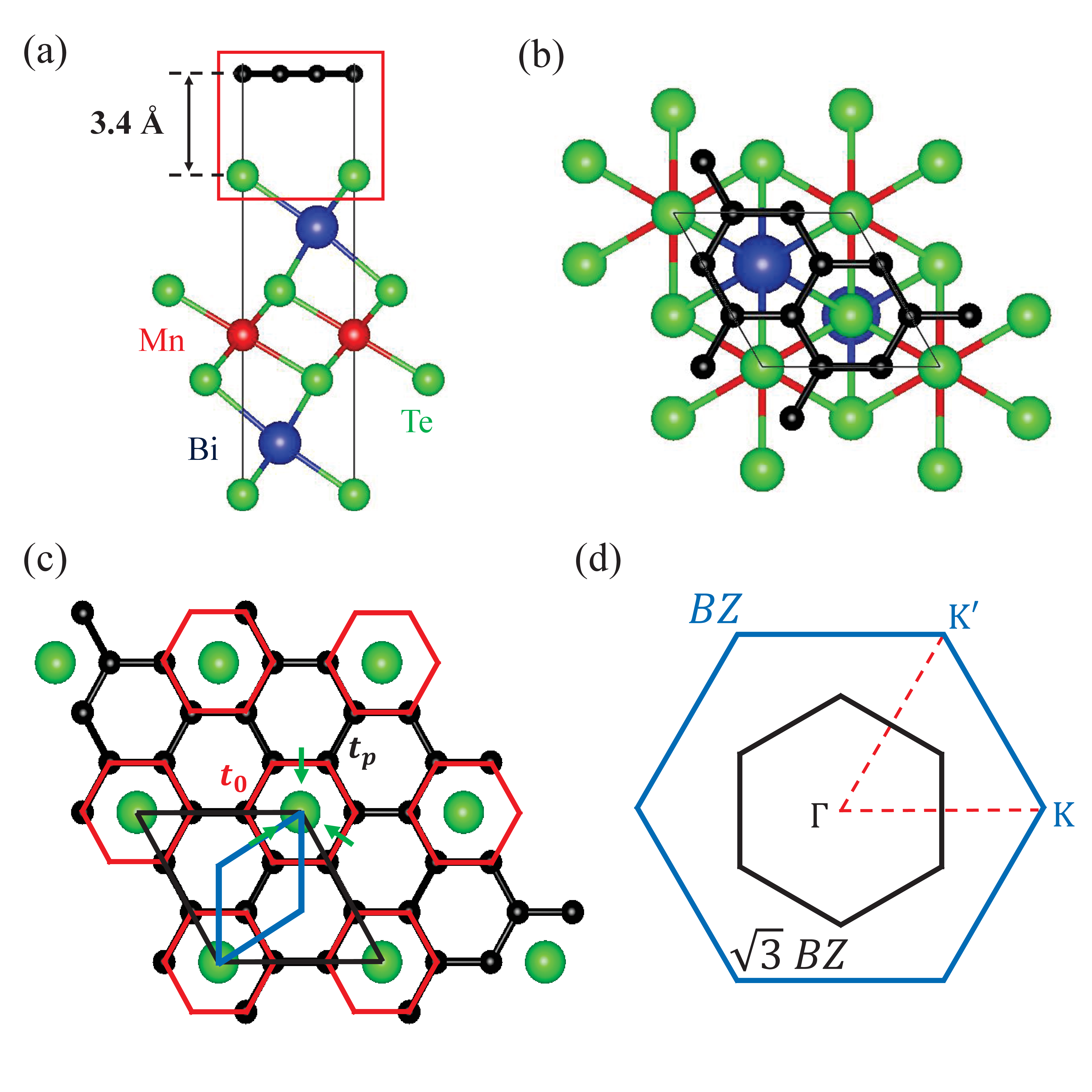}
\caption{The crystal structure of the graphene/MBT heterostructure in (a) side view and (b) top view. The black, red, blue, and green atoms indicate the C, Mn, Bi, Te atoms, respectively. (c) Illustrates Kekul$\rm \acute{e}$ distortion with two distinct hopping strengths: the red bonding involves the hopping surrounding Te atoms with the strength $t_0$, while the black bonding connects the red ring with a hopping strength $t_p$. The blue parallelogram denotes a graphene primitive cell, and the black one is a commensurate $\sqrt{3}\times\sqrt{3}$ graphene supercell. Green arrows indicate the direction of Rashba SOC induced by the in-plane electric field surrounding the Te atoms. (d) Brillouin zone of graphene primitive cell (blue line) and the $\sqrt{3}\times\sqrt{3}$ commensurate cell (black line).}
\label{fig:01}
\end{figure}

 The graphene/MBT heterostructure consists of monolayer graphene and one septuple-layer MBT, as illustrated in Fig.~\ref{fig:01}(a) for the side view and Fig.~\ref{fig:01}(b) for the top view. The in-plane lattice constant of MBT is approximately 4.33$\rm \AA$, while the $\sqrt{3}\times\sqrt{3}$ supercell constant of graphene is around 4.27$\rm \AA$. Consequently, a 1.4\% lattice mismatch exists between MBT and $\sqrt{3}\times\sqrt{3}$ graphene. Due to the mechanical flexibility of graphene, the lattice mismatch, along with the topmost Te layer, locally stretches the graphene lattice, leading to Kekul$\rm \acute{e}$ distortion. The Kekul$\rm \acute{e}$ distortion breaks the bond symmetry in graphene, forming equivalent C-C bonding into two distinct red and black bonds in Fig.~\ref{fig:01}(c) with different hopping strengths\cite{NO24,NO25}. To describe the electronic properties of graphene/MBT heterostructure, we employ an effective graphene supercell Hamiltonian by introducing SOC, Kekul$\rm \acute{e}$ distortion, and Zeeman exchange field in graphene:
\begin{equation}\label{eq:01}
\begin{split}
    H=&\  \sum_{k=0,p}t_k \sum_{\left \langle i,j \right \rangle_{\alpha}} c_{i\alpha}^{\dag}c_{j\alpha} + \frac{i\lambda_I}{3\sqrt{3}}\sum_{\left \langle \left \langle i,j \right \rangle \right \rangle_{\alpha\beta}} \nu_{ij}c_{i\alpha}^{\dag}s^zc_{j\beta}\\
    & + i(\lambda_R^z \hat{\bf{z}} + \lambda_R^{\rho} \boldsymbol{\hat{\rho}}) \sum_{\left \langle i,j \right \rangle_{\alpha\beta}}c_{i\alpha}^{\dag} (\boldsymbol{ s \times \hat{d}}_{ij})_{\alpha \beta} c_{j\beta}\\
    & + i\frac{2}{3}\lambda_{R2}\sum_{\left \langle \left \langle i,j \right \rangle \right \rangle_{\alpha\beta}}\mu_ic_{i\alpha}^{\dag}(\boldsymbol{s\times\hat{D}}_{ij})^z_{\alpha\beta} c_{j\beta}\\
    & + M\sum_{i\alpha}c_{i\alpha}^{\dag}s^zc_{i\alpha}
\end{split}
\end{equation}
\noindent where $c_{i\alpha}^{\dag}$ is the creation operator and $c_{j\alpha}$ is the annihilation operator of an electron with spin polarization $\alpha$ at site $i$. $\left \langle i,j \right \rangle$ and $\left \langle \left \langle i,j \right \rangle \right \rangle$ run over the nearest neighbor and next-nearest neighbor hopping sites with unit vector $\boldsymbol{\hat{d}}_{ij}$ and $\boldsymbol{\hat{D}}_{ij}$, respectively. The first term describes the nearest neighbor hopping of graphene with two different hopping strengths: the $t_0$ with red-bonding carbon rings surrounding the top Te atoms and $t_p$ connecting the rings (Fig.~\ref{fig:01}(c)). The hopping difference reflects the Kekul$\rm \acute{e}$ distortion in graphene supercell, opening a band gap of $2|t_0-t_p|$. The second term represents the intrinsic SOC, which opens a non-trivial band gap of $2\lambda_I$, where $\boldsymbol{s}=(s^x,s^y,s^z)$ is the Pauli matrix in the spin degree of freedom. The parameter $\nu_{ij}$ describes the next-nearest neighbor hopping in a clockwise/counterclockwise direction to the positive z-axis with +1/-1 values. The third and fourth terms account for the Rashba SOC. The third term $\lambda_R^z$  associated with the nearest neighbor hopping, is induced by the external electric field perpendicular to the graphene sheet from the MBT substrate, breaking the out-of-plane inversion symmetry. The fourth term $\lambda_R^{\rho}$ arises from the nonuniform in-plane electric field, represented by the green arrow in Fig.\ref{fig:01}(c), breaking the in-plane inversion symmetry. The fifth term is the second-order Rashba SOC $\lambda_{R2}$, associated with next-nearest-neighbor hopping, with $\mu_i=+1(-1)$ for the A(B) site. The sixth term represents the Zeeman exchange field magnetization originating from the Mn atoms in MBT substrate.

\par We explore the topological phase diagram in eq(1) through the utilization of two fundamental topological invariants, Chern number $C$ and spin Chern number $C_s$. For the Chern insulator, as observed in the QAH effect, the quantized charge Hall conductivity $\sigma_{xy}=C*e^2/h$ is directly proportional to an integer denoted as Chern number $C$. The Chern number can be calculated from the integral of Berry curvature in momentum space and sum over all occupied bands below the bulk band gap, which is directly related to the number of chiral edge states in the two-dimensional topological system\cite{NO3,NO26}. In a $s_z$ conserved system, the spin Chern number $C_s$ is identical to $\mathbb{Z}_2$ index owing to the good quantum number $s_z$. The $C_s$ is expressed as $C_s=\frac{1}{2}(C_\uparrow-C_\downarrow)$, where $C_\uparrow/C_\downarrow$ is computed similarly to the Chern number, but the occupied states are separated by spin-up/spin-down electrons\cite{NO11,NO12}. This configuration characterizes the system as a spin Chern insulator, observed in the QSH effect. Recent studies show that even in scenarios where $s_z$ is not a good quantum number, the $C_s$ remains well-defined through the introduction of Rashba SOC\cite{NO27,NO28,NO29,NO30}. By calculating these two critical invariants, we identify the topological phase at each point within the parameter space of the phase diagram.

\begin{figure}[t]
\centering
\includegraphics[width=\linewidth]{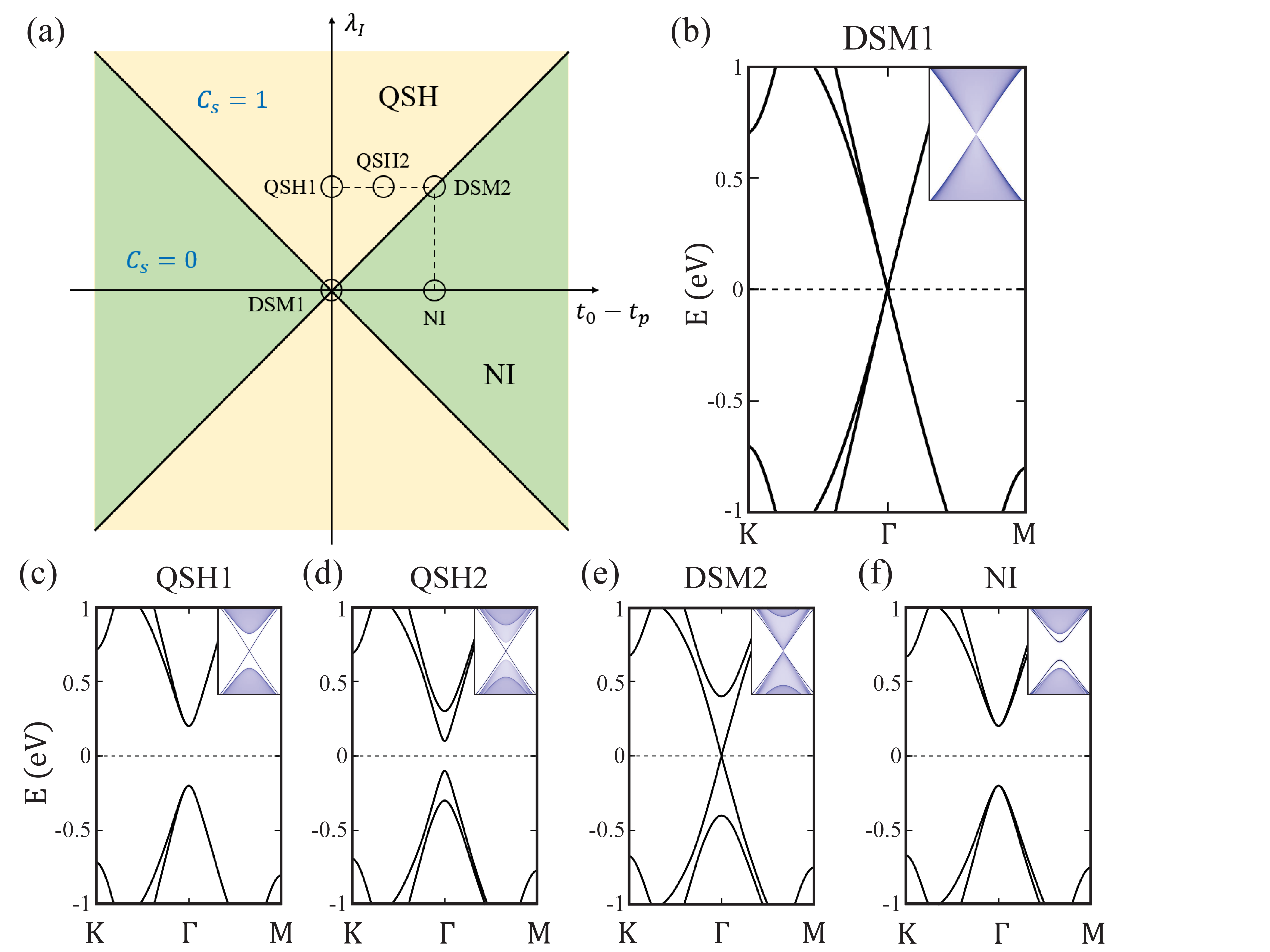}
\caption{(a) The topological phase diagram, with intrinsic SOC (vertical axis) and Kekul$\rm \acute{e}$ distortion (horizontal axis). The heavy lines represent phase transition boundaries, indicating the Dirac semimetal state. The topological invariant in this phase diagram is represented by the spin Chern number $C_s$. The dashed lines indicate the selected phase evolution. (b) The bulk band structure of Dirac semimetal phase (DSM1). (c-d) The spin Chern insulating phase in QSH1 and QSH2, both without and with Kekul$\rm \acute{e}$ distortion, respectively. (e) DSM2 phase, denoting the critical point of phase transition between the spin Chern insulator and normal insulator (NI). (f) NI phase. The inset in (b-f) shows the edge band structure on the armchair edge.}
\label{fig:02}
\end{figure}

\par We initially present the topological phase diagram, delineating the interplay between the two band gap opening mechanisms, Kekul$\rm \acute{e}$ distortion and intrinsic SOC, in the absence of exchange field M (Fig.~\ref{fig:02}(a)). The two Dirac cones within the graphene unit cell distinctly segregate into two valleys, $K$ and $K’$, positioned at the corner of the Brillouin zone (BZ). Considering the periodicity with the MBT substrate, the unit cell of the heterostructure aligns with a commensurate $\sqrt{3}\times\sqrt{3}$ graphene supercell. This alignment folds both Dirac cones from the corner of the original BZ of graphene to $\Gamma$ point in momentum space of the supercell (Fig.~\ref{fig:01}(d)). This folding effect gives rise to the Dirac semimetal DSM1 phase, illustrating an eight-fold degenerate Dirac point at the $\Gamma$ point (Fig.~\ref{fig:02}(b)). Upon the introduction of intrinsic SOC $\lambda_I$, this highly degenerate Dirac state undergoes band gap opening (Fig.~\ref{fig:02}(c)). Due to the spin and valley degeneracy, the valence and conduction bands that separate by a gap become four-fold degenerate at the $\Gamma$ point. In this scenario, graphene transforms from DSM1 to a spin Chern insulator with $C_s=1$ (labeled QSH1). The boundary state with the armchair edge showcases a non-trivial edge state connecting the valance and conduction bands, characteristic of the QSH effect (inset of Fig.~\ref{fig:02}(c)). In contrast to the intrinsic SOC, Kekul$\rm \acute{e}$ distortion introduces a superlattice potential in the graphene supercell, resulting in the opening of a band gap for the Dirac cones with $C_s=0$. The edge state calculations reveal a gapped edge state, as depicted in the inset of Fig.~\ref{fig:02}(f). Due to the competing mechanisms of the intrinsic SOC and Kekul$\rm \acute{e}$ distortion, the ground state of graphene supercell will lie between QSH1 and NI. For instance, Fig.~\ref{fig:02}(d) illustrates a spin Chern insulator induced by the cooperative action of intrinsic SOC and Kekul$\rm \acute{e}$ distortion (labeled QSH2). The doubly degenerated band dispersion arises from the PT-symmetry, wherein opposite spin states degenerate in different valleys. In this case, $\lambda_I>|t_0-t_p|$, thus the stronger intrinsic SOC results in the spin Chern insulator phase with the non-trivial edge state (inset of Fig.~\ref{fig:02}(d)). When $\lambda_I=|t_0-t_p|$, a topological phase transition critical point, a Dirac semimetal phase with four-fold degenerate Dirac point (DSM2) emerges between QSH2 and NI, as presented in Fig.~\ref{fig:02}(e).

\begin{figure}[t]
\centering
\includegraphics[width=\linewidth]{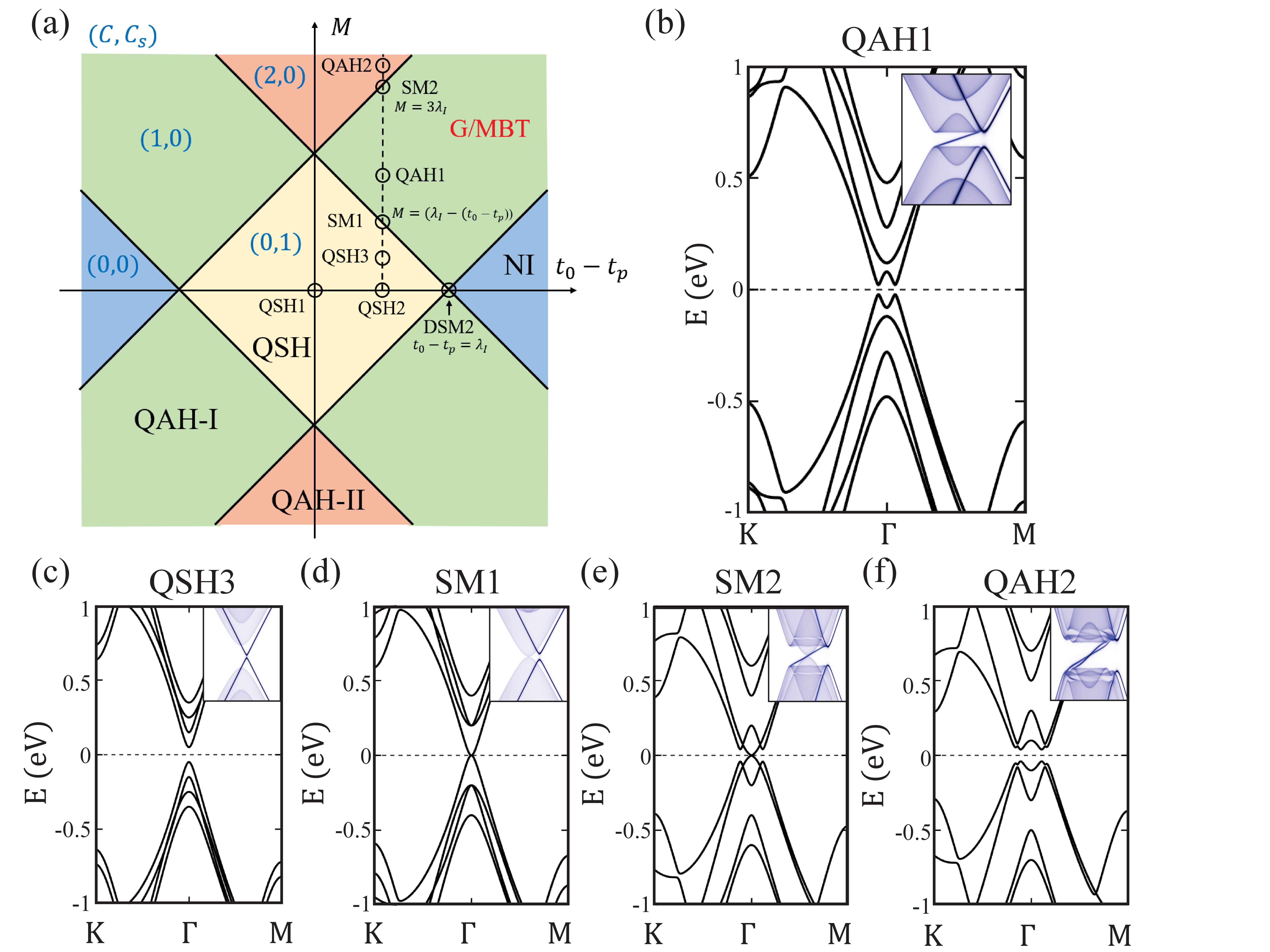}
\caption{(a) The topological phase diagram, with  Kekul$\rm \acute{e}$ distortion (horizontal axis) and Zeeman exchange field (vertical axis). Here the intrinsic SOC and Rashba SOC are included. The heavy lines represent phase transition boundaries, indicating the semimetal state. The topological invariant in this phase diagram is represented by the spin Chern number $C_s$ and Chern number $C$. The dashed lines indicate the selected phase evolution. (b) Chern insulating phase (QAH1). (c) Spin Chern insulating phase with finite Zeeman exchange field (QSH3). (d) SM1 phase, denoting the critical point of phase transition between QSH3 and QAH1. (e) SM2 phase, denoting the critical point of phase transition between QAH1 and QAH2. (f) Chern insulating phase with $C=2$ phase (QAH2). The inset in (b-f) shows the edge band structure on the armchair edge.}
\label{fig:03}
\end{figure}

\par Subsequently, we extend our analysis to include the Zeeman exchange field $M$ in graphene supercell, simulating the graphene/MBT heterostructure. The intricate interplay among SOC, Kekul$\rm \acute{e}$ distortion, and the Zeeman effect give rise to a comprehensive phase diagram. For example, commencing with the QSH2 characterized by a band gap $E_{g(QSH2)}=\lambda_I-(t_0-t_p)$ (Fig.~\ref{fig:02}(d)), the introduction of the exchange field $M$ induces a time-reversal breaking spin Chern insulator (labeled QSH3) when $M<E_{g(QSH2)}$. The topological edge state of QSH3 shifts away from the $\Gamma$ point due to the breaking of time-reversal symmetry. Simultaneously, this edge state opens a gap when further considering the effect of Rashba SOC\cite{NO27}. At a critical point, $M=E_{g(QSH2)}$, a two-fold degenerated quadratic band dispersion emerges (labeled SM1, Fig.~\ref{fig:03}(d)). Subsequently, a band inversion occurs in the parameter region of $E_{g(QSH2)}<M<3E_{g(QSH2)}$, leading the system into a Chern insulator with $C=1$ (labeled QAH1, Fig.~\ref{fig:03}(b)). The inset of Fig.~\ref{fig:03}(b) displays a chiral edge state of QAH1. With a further increase in the strength of the Zeeman exchange field, the evolution of the band structure progresses through another two-fold quadratic semimetal phase SM2 (Fig.~\ref{fig:03}(e)), and then a second band inversion transpires, resulting in a Chern insulator with $C=2$ (labeled QAH2, Fig.~\ref{fig:03}(f)). The inset of Fig.~\ref{fig:03}(f) exhibits two chiral edge states within the bulk gap of QAH2.

\par
\begin{figure}[t]
\centering
\includegraphics[width=\linewidth]{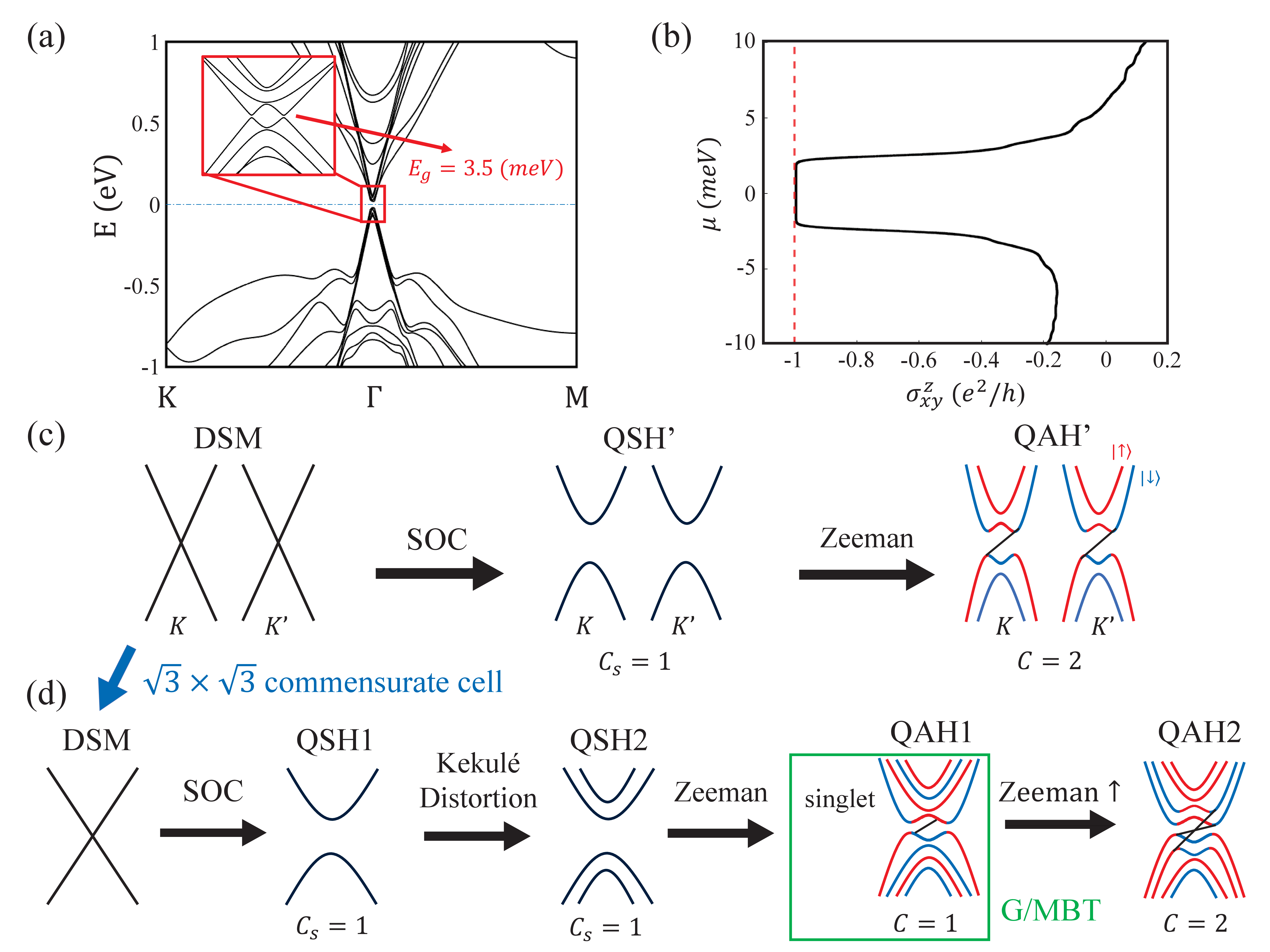}
\caption{(a) Band structure of graphene/MBT heterostructure calculated by DFT. The zoom-in band structure around $\Gamma$ point is shown in the inset. (b) The Hall conductivity was calculated using the Kubo formula. (c-d) Schematic representation illustrates the formation of the Chern insulating state in (c) graphene unit cell and (d) $\sqrt{3}\times\sqrt{3}$ commensurate supercell. The green square indicates the G/MBT heterostructure phase.}
\label{fig:04}
\end{figure}

 To pinpoint the ground state of graphene/MBT heterostructure within the presented parameter space of the phase diagram (Fig.~\ref{fig:03}(a)), we conducted the first-principles calculations using DFT. The calculated band structure of the graphene/MBT heterostructure is depicted in Fig.~\ref{fig:04}(a). It is noteworthy that each band exhibits singly degenerate with a band gap of 3.5 meV. Additionally, the zoom-in band structure reveals a resemblance to band dispersion observed in QAH1 (Fig.~\ref{fig:03}(b)). The Hall conductivity calculated using the Kubo formula showcases a quantized value of $e^2/h$, indicative of Chern insulator with $C=1$ (Fig.~\ref{fig:04}(b)). This finding starkly contrasts with previous investigations where the Chern number was identified as $C=2$ (labeled QAH’ in Fig.~\ref{fig:04}(c)). The QAH state in earlier studies result from the band inversion at both valleys, induced by the interplay of SOC and Zeeman exchange field. This mechanism, though conceptually straightforward, lacks consideration for substrate effects, making experimental realization challenging. Fig.~\ref{fig:04}(d) provides a schematic representation systematically illustrating the formation of the Chern insulating state with $C=1$ in graphene/MBT heterostructure. The original unit cell of graphene harbors two Dirac cones at $K/K’$ valleys. These Dirac cones fold to $\Gamma$ point, forming the DSM phase within the context of a $\sqrt{3}\times\sqrt{3}$ commensurate supercell of graphene. The introduction of SOC opens band gap at both $K/K’$ valleys, giving rise to the QSH1.  Upon further consideration of substrate effect, specifically Kekul$\rm \acute{e}$ distortion, QSH1 transitions into QSH2. In this phase, two valleys at the original BZ of graphene become intertwined exhibiting distinct energy eigenvalues. Consequently, the Zeeman exchange field selectively inverts the topmost valence band and bottommost conduction band, leading to the emergence of QAH1 with $C=1$. 

\section{\label{sec:level1}Discussion and Conclusion}

\par Our systematic study unveils the capability of the graphene/MBT heterostructure to control topological phase transitions, offering avenues for manipulation through methods, such as doping or stacking additional van der Waals layers to adjust the Fermi level and/or the strength of exchange coupling\cite{NO31,NO32,NO33}. For instance, the Zeeman exchange field can be enhanced through increasing the stack of MBT layers and MnTe layers\cite{NO31} to produce the QAH2 $C=2$ phase in Fig.~\ref{fig:03}(f). These advantages present a tunable platform for quantum Hall devices and advancing spintronics.

\par In summary, our work systematically investigated the realization of the QAH effect on monolayer graphene through an effective model Hamiltonian and first-principle calculations. We discern that the graphene/MBT heterostructure manifests a Chern insulating phase with $C=1$, considering the interplay of the SOC, Kekul$\rm \acute{e}$ distortion, and Zeeman exchange interaction within a fruitful topological phase diagram. Leveraging the flexibility inherent in both MBT families and graphene systems, coupled with the previous established experimental advantage of graphene/$\rm Bi_2Te_3$ systems\cite{NO22}, our work not only introduces a novel avenue for achieving the QAH effect in monolayer graphene but also delves into magnetic topological phase transition within 2D heterostructure.

\section{\label{sec:level1}Method}

\par The graphene/MBT first-principle calculations are presented by density-functional theory (DFT) with the projector augmented wave method implemented in the Vienna ab-initio simulation package (VASP)\cite{NO34,NO35,NO36}. The top and bottom vacuum space is 10$\AA$ to avoid the interaction between two adjacent heterostructures. The structure in Fig.~\ref{fig:01}(a) is fully relaxed until force on each atom is smaller than 0.01 eV/$\AA$. The interlayer distance between graphene and MBT is 3.4$\AA$. We used the Monkhorst-pack grides with 11×11×1 mesh. The exchange-correlation method with GGA+U is adopted with Hubbard U is 4 eV. The Heyd-Scuseria-Ernzerhof (HSE) hybrid function\cite{NO37} is used to correct the band gap of monolayer MBT. The van der Wall interaction correction using the Grimme (DFT-D2) method\cite{NO38} was considered.

%

\end{document}